\documentclass[runningheads]{svmult}

\usepackage{graphicx}  
\usepackage{subeqnar}  
\usepackage{multicol}  
\usepackage{physprbb}  

\begin{document}
\title*{Models for Type Ia Supernovae and Cosmology}
\toctitle{Models for Type Ia Supernovae
\protect\newline  and Cosmology}
\titlerunning{Type Ia Supernovae and Cosmology}
\author{Peter H\"oflich \inst{1},
Chris Gerardy \inst{1}, 
Eric Linder \inst{2}, Howie Marion \inst{1}}
\authorrunning{Peter H\"oflich et al.}
\institute{Dept. of Astronomy, University of Texas, Austin, TX 78681, USA
\and Lawrence Berkeley Laboratory,  Berkeley, CA 94720, USA}

\maketitle

\begin{abstract}
 We give an overview of the current understanding of Type Ia supernovae
relevant for their use as cosmological distance indicators. We present the physical
basis to understand their homogeneity of the observed light curves and spectra and
the observed correlations. This provides a robust method to determine the Hubble constant,
  $67 \pm 8 (2 \sigma)  km~Mpc^{-1}~sec^{-1}$,
independently from primary  distance indicators.
 We discuss the uncertainties and tests which include SNe~Ia based distance determinations
 prior to $\delta $-Ceph measurements for the host galaxies. Based on detailed models,
 we study the small variations from homogeneities and their observable consequences. 
In combination with future data,
this  underlines the suitability  and  promises the refinements needed to determine accurate
relative distances within 2 to 3 \% and
to use
SNe~Ia for high precision cosmology.
\end{abstract}

\section {Overview} 

Type Ia Supernovae (SNe~Ia) are the result of a thermonuclear explosion of a white dwarf
star. What we observe is not the explosion itself but light emitted from the 
material of the disrupted white dwarf (WD) for weeks to months afterward. After the
first few seconds, this rapidly moving gas expands freely. As a consequence, the 
matter density  decreases with  time and the expanding material becomes 
increasingly transparent, allowing us to see progressively deeper layers. Thus, 
a 
detailed analysis of the observed light curves (the time series of  emitted 
flux) and
spectra reveals the density and chemical structure of the entire star. 

The 
structure of a WD is determined by degenerate electrons and thus 
largely independent of details such as the temperature or chemical composition. 
The explosion energy is determined by the binding energy released during the 
nuclear burning, and the burning products can be observed. The tight relation 
between the explosion and the observables and their insensitivity to details are  
the building blocks on which our understanding of the homogeneity in the  
observable relations for SNe~Ia is based. Indeed, both the 
peak fluxes and light curve shapes of SNe Ia 
show an impressive level of homogeneity, making them the astronomical objects 
closest to a standard candle distance estimator.  This allows their use for 
precision estimation of  cosmological parameters. 

The thumbnail sketch of our 
understanding is as follows: 
\begin{itemize} 
\item Type Ia supernovae are nearly 
homogeneous because  nuclear physics determines the structure of white 
dwarfs, and the explosion. 

\item The total production of nuclear energy is 
almost constant since very little of the WD remains unburned. The final 
explosion 
energy depends on the binding  energy of the WD, which is given by its 
structure.

\item The light curves are powered by the radioactive decay of $^{56}Ni$ 
produced
during the explosion, independently from details of the explosion physics and 
progenitors. The amount of $^{56} Ni$ determines the
absolute
brightness. 

\item The energy released from the nickel decay ties together the 
luminosity and the temperature dependent opacity, i.e.~how much flux is emitted 
and how quickly. Explicitly, less Ni means a lower luminosity, but at the same 
time lower temperature in the gas and so lower opacity. Thus, energy escape is
more rapid. So dimmer SN are quicker, i.e.~have narrower light curves. 
 This is variously called the brightness decline, peak magnitude –- light curve 
width, or
stretch relation.

\item To 
be in agreement with the narrowness of the brightness decline relation 
\cite{phillips87},
 the mass 
of the progenitors and the explosion energies must be similar. This is 
automatically satisfied in the currently most successful model:  a
Chandrasekhar mass C/O-WD in which the burning  starts off as
a deflagration front (propagating at well below the speed of sound) and
subsequently turns into a detonation (with $\approx $ the speed of sound).

\item To agree with observations of intermediate mass elements at the outer 
layers, the WD must be pre-expanded. Most likely, an initial deflagration phase 
causes the pre-expansion. This depends mainly  on the amount of energy release
but 
not on the details of the deflagration front. Within this paradigm, 1) the 
entire
WD is burned, and 2) the production of $^{56}Ni$ is dominated by a single
parameter characterizing  the transition between deflagration and detonation,
determining the amount of burning during the deflagration. 

\indent 
~~~The deflagration-detonation model thus gives a natural and well motivated
origin 
for  a narrow brightness decline relation. In addition, the resulting chemical 
layering is shell like as observed. 

\item Homogeneity can be established down to 
a level of 0.2 
magnitudes. Beyond this, secondary parameters are expected to become important, 
namely the progenitor mass on the main sequence, its metallicity, and stellar 
rotation. In particular, the pre-conditioning of the WD prior to the 
thermonuclear runaway may hold the key to understanding the variety of SNe~Ia. 

\indent 
~~~With more detailed observations, these characteristics 
will help to improve the current accuracy of SNe~Ia as distance 
indicators. 

\end{itemize}  

In the following sections, we address the current 
status of our understanding of SNe in more detail and elaborate
on how future observations by ground based telescopes and dedicated space 
missions, in 
combination with detailed modeling, will help to bring us to a new level of 
understanding. These include new insights into the nature of SNe
including 
the progenitors, the thermonuclear runaway that leads to the explosion, the 
propagation of nuclear burning fronts and their 3-dimensional nature. These 
studies will help to discover and understand new relations between observables 
to get a handle on the relation of SNe~Ia with their environment, including 
evolutionary effects with redshift, and to improve the accuracy of SNe~Ia as 
cosmological distance indicators.

\section{A Simplified Explosion Model} 

The 
following scenario summarizes one possibility for the creation and 
characteristics of a SNe~Ia. This is designed solely to give the
reader a simple example to relate a variety of concepts.

\begin{figure}[!b]
\hskip 0.3cm \includegraphics[width=8.4cm,angle=270]{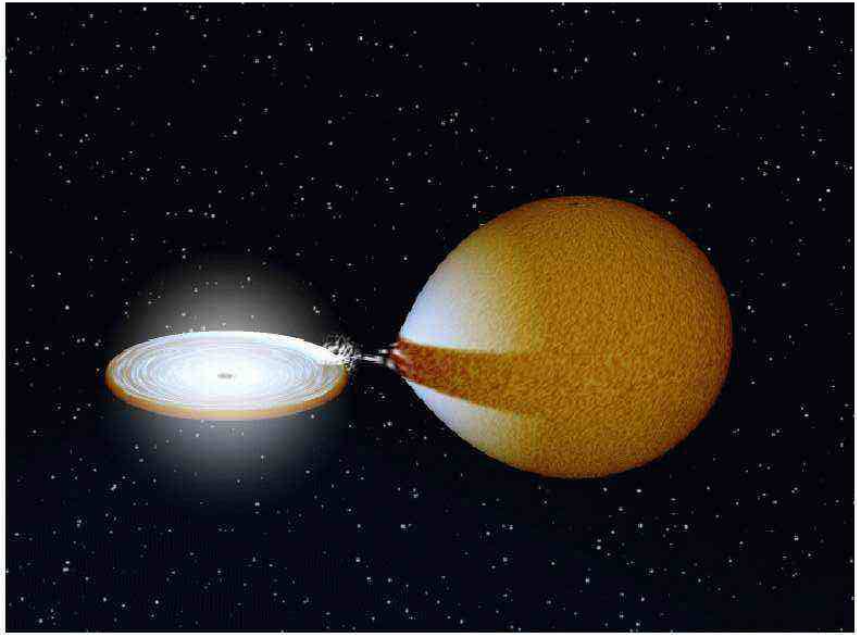}
\vskip -0.0cm
\caption{Possible scenario for a 
progenitor system of a SN~Ia. A white dwarf accretes material from
a 
close companion by Roche lobe overflow. Initially, the WD has a mass between 0.6 
and 1.2 $M_\odot$ and, by accretion, approaches the Chandrasekhar mass limit. 
The
companion star may be a main sequence star or red giant, or a  helium star
or another WD. Depending on this, the accreted material may be either H, He or 
C/O rich. If H or He is accreted, nuclear burning on the surface converts it to 
a C/O
mixture at an equal ratio in all cases.
 Despite the different evolutionary pathways, the final result will be the same: 
the
explosion of a C/O-WD with a mass close to $M_{Ch}$ and with very similar SN
properties. Some small fraction of SNe~Ia may also be the result of merging of 
two WDs on a dynamical time  scale.} 
\label{bin} 
\end{figure} 

Consider a white dwarf (WD) in a binary system, accreting mass from its companion (\ref{bin}).
  This initial phase
ends when the total mass approaches the Chandrasekhar mass (beyond which the 
star would collapse to a neutron star or black hole) and causes 
compressional heating of the core and the thermonuclear runaway.  Likely,
the burning front starts as a deflagration (velocity well below the sound speed).
 The energy release lifts the WD in its potential
and causes pre-expansion of the star needed to reduce the density under which burning occurs.
 After a few seconds, the burning fronts makes a transition to a detonation (or very fast deflagration).
 All material in the high density regions is burned to $^{56}Ni$ while outer shells 
of Si, S, etc. and a small fraction of the original C and O remain. The specific energy released by 
these reactions unbinds the material and causes a rapid acceleration of the matter.
 As the radioactive $^{56}Ni$ decays, the resulting gamma ray energy is thermalized and 
produces the optical luminosity, or light curve (LC), over tens of days to months.
The rise  time and decay time are given by the decay rate and the opacity and expansion.
 Spectral time series map out the SN structure  as the
photosphere recedes through the material. 

 The simplified physics picture is as follows:
 The rate of the free expansion
is  determined by the specific nuclear energy production which, for a C/O-WD, is
rather insensitive to the burning process and the final burning products.
 The complete burning of the white dwarf in the explosion
fixes the total nuclear energy release and hence kinetics; the transition
density determines the nickel mass produced; the nickel decay fixes the energy input to
the supernova material, determining its luminosity and opacity. The opacity and explosion energy 
together give the shape of the light curve, namely the brightness decline relation.

Alternate model pathways, in fact, converge to the same
major points after each stage, driven by the physics and 
constrained by the observations. Several examples of such ``stellar amnesia''
indemnify the observables against details of how the final state is reached. 
 For example what is predominantly 
important for the energy production is that the overwhelming majority of the 
star burns, not how it does because the release of energy by the fusion of C/O
up to Si/S dominates over the relatively little binding energy in
the last stage to iron.  

    On the other hand, the tight relation between the observables evinced by the 
homogeneity of the brightness-width relation only occurs in certain classes of 
models.  This constrains the possibilities, as do such data as infrared spectra 
showing little unburned original C-O material.  Together, amnesia from the 
physics and empirical data from observations weave a tight net around the 
possible ingredients that can be important in determining the 
absolute brightness.

\section{Physics of the Explosion, Light Curves, and Spectra}

As stated in the overview, nuclear physics determines the structure of the progenitor and
 is responsible for the homogeneity of SNe~Ia.
 As the SN~Ia expands, we see deeper layers with time due to the geometrical dilution. 
The unveiling of the layers reveals the structure of the 
WD.  The observable data provide a rich resource for 
testing and refining models.  E.g., the light curves provide
critical information about integrated quantities such as the total energy 
generation and mass and energetics of the expanding material. The spectra are 
mostly sensitive to the composition and velocity at the photosphere (the deepest 
unveiled layer).

The spectra of SNe~Ia are dominated by 
elements (C,O,Si,S,Ca,Fe/Co/Ni) that are the characteristic   products of 
explosive nuclear burning  at densities between $10^{6-9} g/cm^3$. These 
densities are typical only for a C/O-WD, i.e. a star stabilized by a
degenerate electron gas. 
For such a degenerate equation of state, the 
initial structure of the exploding WD depends only weakly on the 
temperature and the C/O ratio as a function of depth. WD radii are between 1500 to 2000 km/sec,
mainly depending on density at the time of the accretion which is mainly
given by the accretion rate (see  \cite{hwt98} and above).
 Typical binding energies are $\approx 5$ to $6 \times 10^{50} erg$.
 If the entire WD is burned, about $2 \times 10^{51} ergs$ are released over time scales of seconds.
  The energy released is given by the
difference between the binding energy per nuclear of unburned compared to burned matter. Because the nuclear binding energy of both the
of the fuel, i.e. carbon and oxygen, and the final burning product, i.e. Si and Ni, are rather similar, variations in the specific energy
release per mass of burned matter is limited to $\approx 10 \%$. Neutrino losses are less than 1 to 2 \% and, thus, little energy is lost in
contrast to core collapse SNe where more $\approx 99 \%$ of the release energy is lost by neutrinos.
 In the explosion, a WD with a radius of about 1500 km expands with observed velocities of the order of 10,000 
km/sec, consistent with the specific nuclear energy release in such 
an environment. Because this rapid increase in volume and the adiabatic cooling, the nuclear energy is used to overcome the binding energy
and to accelerate the WD matter.
  Based on this evidence, there is general agreement that SNe~Ia 
result from some  process of combustion of a degenerate WD.
The amount and products of the nuclear reactions -- ``burning'' -- depend 
mainly on the time scale of reactions compared to the hydrodynamical time 
scale of expansion, which is $\approx 1 sec $.
 The reaction rate depends sensitively on the temperature and
the energy release per volume element.  The specific energy release 
is a function of the density and, to a smaller extent, the initial 
chemical composition, namely the C/O ratio of the progenitor (which 
depends on the initial stellar mass).  At densities $\ge 10^7$, 
$\ge 4\times 10^6$, and $\ge 10^6 g/cm^3$, the main burning products 
are Fe/Co/Ni, S/Si and Mg/O, respectively.  We will see, however, that 
the details of the burning process have little 
impact. These quantities we discussed -- the explosion energy, mass, and the 
burning product -- are directly linked in SNe~Ia, and  accessible to 
observations. 

As just mentioned,
virtually none of the initial stored energy from the WD will contribute to the
luminosity of the supernova  but it goes to expansion.
 Instead, the energy input is entirely caused by 
the radioactive decay of freshly synthesized $^{56}Ni$ that decays via $ ^{56}Ni 
\rightarrow ^{56}Co \rightarrow ^{56}Fe$ with life times of 8.8 and $\approx 
111$ days, respectively. This  slow nuclear energy release is due to a gain of
nuclear binding energy between isotopes  rather than change of elements. The total energy
released by radioactive decays is about 3 \% of the initial energy release, namely,
 $\approx 7 \times 10^{49} erg$ for a $^{56}Ni$ production of 0.5 $M_\odot$    vs. $2 \times 10^{51} erg$ released during
the early explosive burning and any changes in the expansion velocities are less than $2 \%$.
The energy release in the form of luminosity is dominated by the location of the 
photosphere within the expanding material, a function of the expansion and the opacity. 
 As we will see below, small differences in the expansion rate caused by the central density of the WD,
chemistry and expansion rate will produce small deviations from the homogeneity in the light curves on a 10 to 20 \% level.

In the case of a Type Ia SN, acceleration of the material takes 
place during the  first few seconds to minutes, followed by the phase of 
free expansion. In 
lack of further acceleration, the radius of a gas element from the
center is simply proportional to the velocity $r\sim v$. This means 
that gas further out moves faster and hence stays further out; each 
shell expands without crossing another, maintaining the original 
structure.  As in cosmology
it is useful to think in comoving coordinates, moving with the expansion.
In these coordinates each shell, and hence slice of the original stellar 
structure, is preserved.  So while material expands out through a fixed 
radial distance from the center, the mass within a comoving radius is 
constant with time. Therefore we often discuss the structure in terms 
of mass or velocity coordinates. 
Due to the expansion, the material cools almost adiabatically 
as the volume increases rapidly:  $V\sim r^3\sim t^3$.  The increase
in volume causes a corresponding decrease in density and hence optical depth. 
So the photosphere slips deeper within the material, simultaneously 
allowing us to see further in.  Note though that it still expands in 
physical radius, at least up to the time of peak magnitude.

The well determined energy source for the luminosity and the tight relation 
between the explosion and the observables, with their simultaneous insensitivity 
to details, are  the building blocks on which our understanding of the 
homogeneity in the  observable relations for SNe~Ia is based. 

To go beyond the homogeneity and take full advantage of the intrinsic 
properties of SNe~Ia and to test 
for the influence of the metallicities, progenitors etc., we can 
perform detailed calculations which are consistent with respect to the 
progenitors, explosion, light curves and spectra.  These calculations 
include detailed nuclear networks, gamma-ray transport, 
non-LTE level populations, and multidimensionality for parts of the 
problem.  The numerical methods are briefly described in the Appendix.
  For more details, see 
\cite{hoeflich93,hoeflich95,hgfs02,h02}, and references
therein.  The only remaining free parameters to address are the initial 
structure of the WD and the description of the nuclear burning
front, which we discuss in the next section.  
\begin{figure}[!h] 
\includegraphics[width=4.5cm,angle=270]{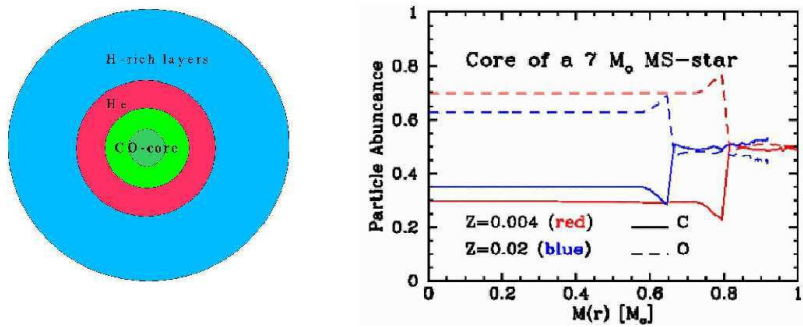}
\vskip -0.0cm
\caption{Stellar structure at the final stage of the evolution of a 
7 $M_\odot$ main sequence star. At this stage, the star loses its H and He
rich 
material with the central C/O-core remaining. This forms the C/O white dwarf
which, eventually, becomes the accreting progenitor in typical SNe~Ia scenarios. 
On the right,
we  show the composition of the core as a function of mass (in
$M_\odot$). The region of reduced C abundance is produced during
the central helium burning which is convective (see also dark green center, left plot).
 Because the size of the helium
burning core  depends on the main sequence mass and the convection depends on the
metallicity, the final structure depends on both (from \cite{hoeflich00}).}
\label{nom}
\end{figure} 

\section{Detailed Models, Observations and Cosmology}

In examining possible scenarios, from the progenitor 
state to the explosion, we will see that the most important properties are 
those that change the overall energetics, such as the total energy 
content of the fuel and the amount of matter of the WD that undergoes 
burning.  As we have alluded to in the previous discussions of "stellar 
amnesia", many of the results are quite stable, i.e.~model independent. 
In fact, we find 1) insensitivity of the WD structure to the progenitor 
star and system.  This is caused by the electron degeneracy enforcing 
the mentioned weak dependence on the temperature and composition.  
2) The time of the explosion (and therefore the WD density) is governed 
by the accretion rate shortly before the thermonuclear runaway causing 
the explosion.  3) Moreover, the final outcome of the explosion is 
rather insensitive to details of the nuclear burning.  While this is 
beneficial for the tightness of the observed luminosity relation, it 
makes it difficult to investigate the pre-supernova physics, e.g.~the 
explosion scenario. 
\begin{figure}[!h] 
\includegraphics[width=5.7cm,angle=270]{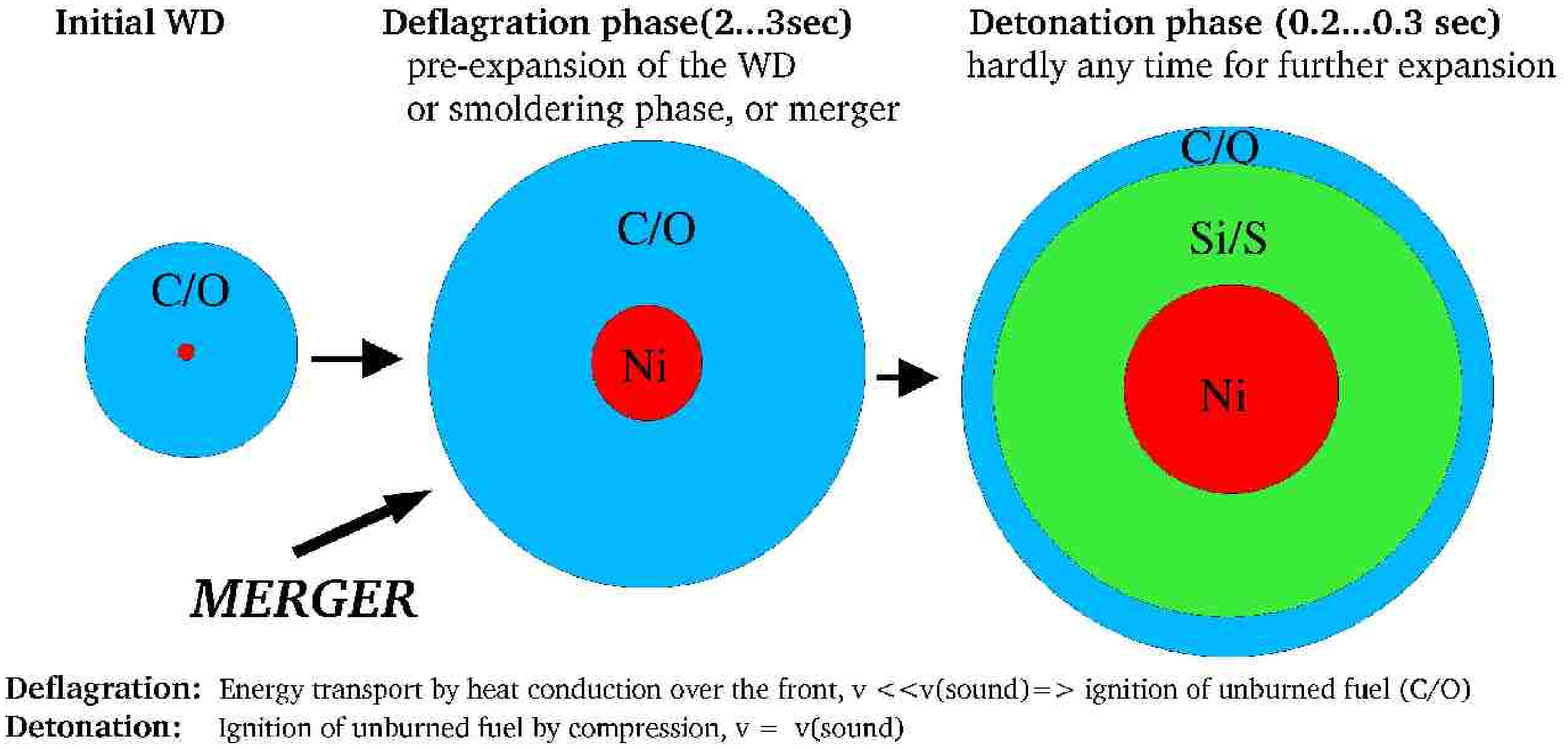}
\vskip -0.0cm
\caption{Schematics of the explosion of a white
dwarf near the Chandrasekhar mass. A thermonuclear runaway occurs near the
center and a burning front propagates outwards (\cite{hs02}). Initially, the burning front
must start as a deflagration to allow a pre-expansion because, otherwise, the 
entire WD would be burned to Ni. Alternatively, the pre-expansion may be 
achieved during the non-explosive burning phase just prior to the thermonuclear 
runaway \cite{hgfs02}.
Subsequent burning is either a fast deflagration or a detonation.  In pure 
deflagration models, a significant amount of matter remains unburned at the 
outer layers, and the inner layers show a mixture of burned and unburned 
material.  In
contrast, the models making a transition to a detonation produce the observed 
layered
chemical structure with little unburned matter, wiping out the history of
deflagration (see text and Fig. \ref{defl}). Note that all scenarios have a
similar, pre-expanded WD as an intermediate state.} 
\label{sce} 
\end{figure} 
\begin{figure}[!h] 
\hskip -0.1cm
\includegraphics[width=4.6cm,angle=270]{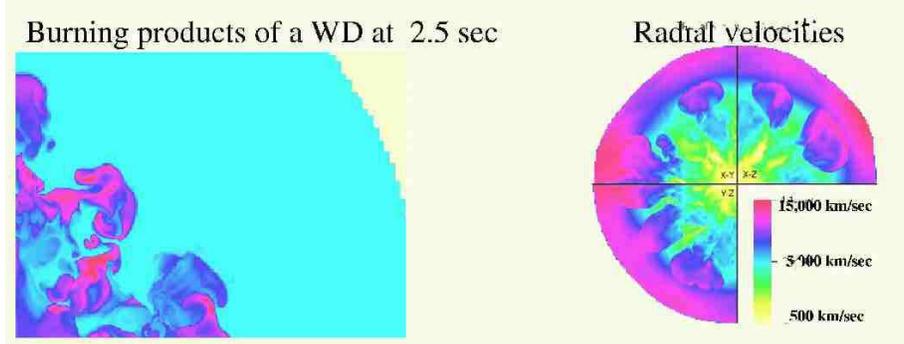}
\vskip -0.0cm
\caption{ Structure of the deflagration front in an 
exploding C/O-WD (left) and the velocity field (right) at about 2
seconds
after runaway based on 3-D calculations by Khokhlov
\cite{khokhlov01}. Light green and  dark red/blue mark 
unburned
and burned
material, respectively. During this phase, the expansion of the material is 
already almost spherical (right), and deviations of the density from sphericity 
are less than 2\%. In normal bright SNe~Ia,  the transition to the
detonation should occur in delayed detonation models at about the time of these
snapshots. In this case, the density in the inner region is sufficient to burn
the unburned material up to Ni, eliminating the chemical contrast and leaving a
layered structure. In contrast, in pure deflagration models, the density will 
drop further before burning can take place, thus leaving intact the chemical 
contrast of the inner layers. In addition, the expansion of the outer layers is 
already close to the speed of sound, faster than the burning front.
 As a consequence, all deflagration models 
show a massive outer layer of unburned  matter.} 
\label{defl}
\end{figure} 

\begin{figure}[!h] 
\includegraphics[width=7.9cm,angle=270]{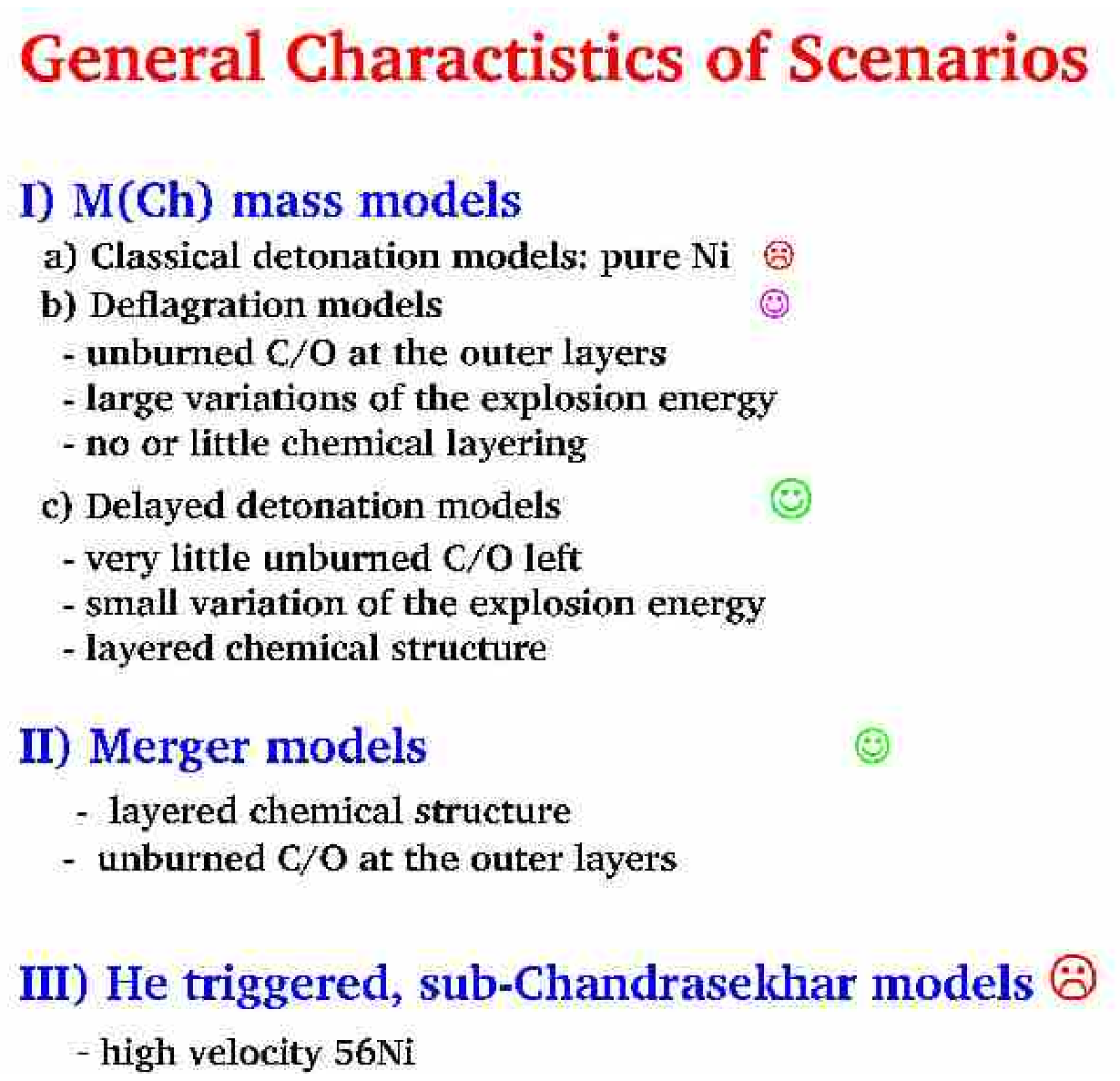}
\vskip -0.0cm
\caption{General properties of various explosion 
scenarios. Delayed detonation models and, possibly, merger models are the 
scenarios  most
likely realized in SNe~Ia. Merger models may contribute to the populations but 
their large amount of unburned C/O at the outer layers is inconsistent with the 
(few) IR-spectra  obtained up to now and can likely constitute only a small 
fraction of the SNe~Ia population. Currently, pure deflagration models show no 
layered
chemical structure, in disagreement
with observations. However,  as of now, 3-D deflagration models consider only 
the
regime of large scale instabilities, i.e.  Rayleigh-Taylor, and start from 
static WDs (see text).}
\label{smiley}
\end{figure} 
\begin{figure}[!t]
\includegraphics[width=5.0cm,angle=360]{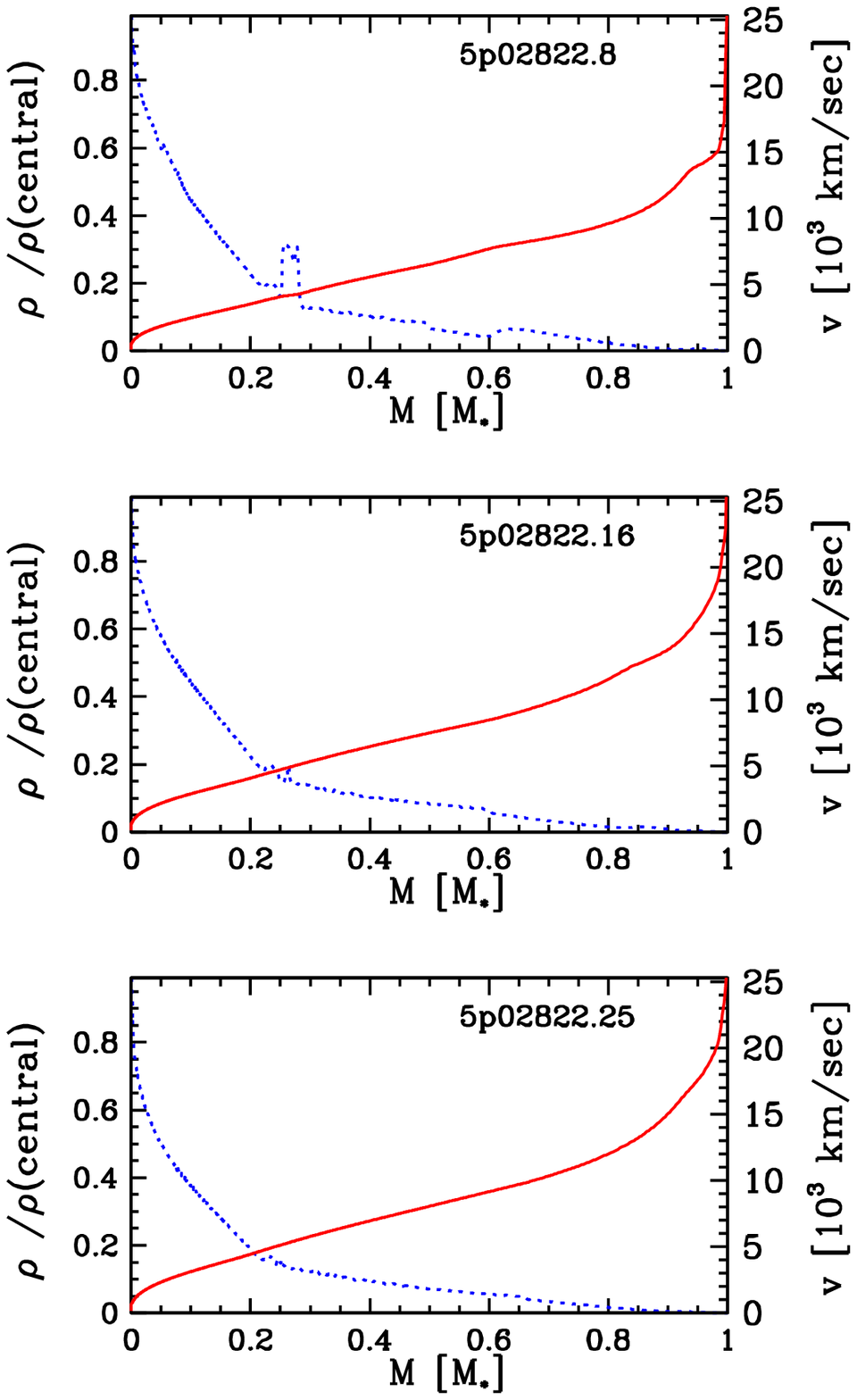}
\vskip -9.0cm
\hskip +6cm
\includegraphics[width=12.0cm,angle=180]{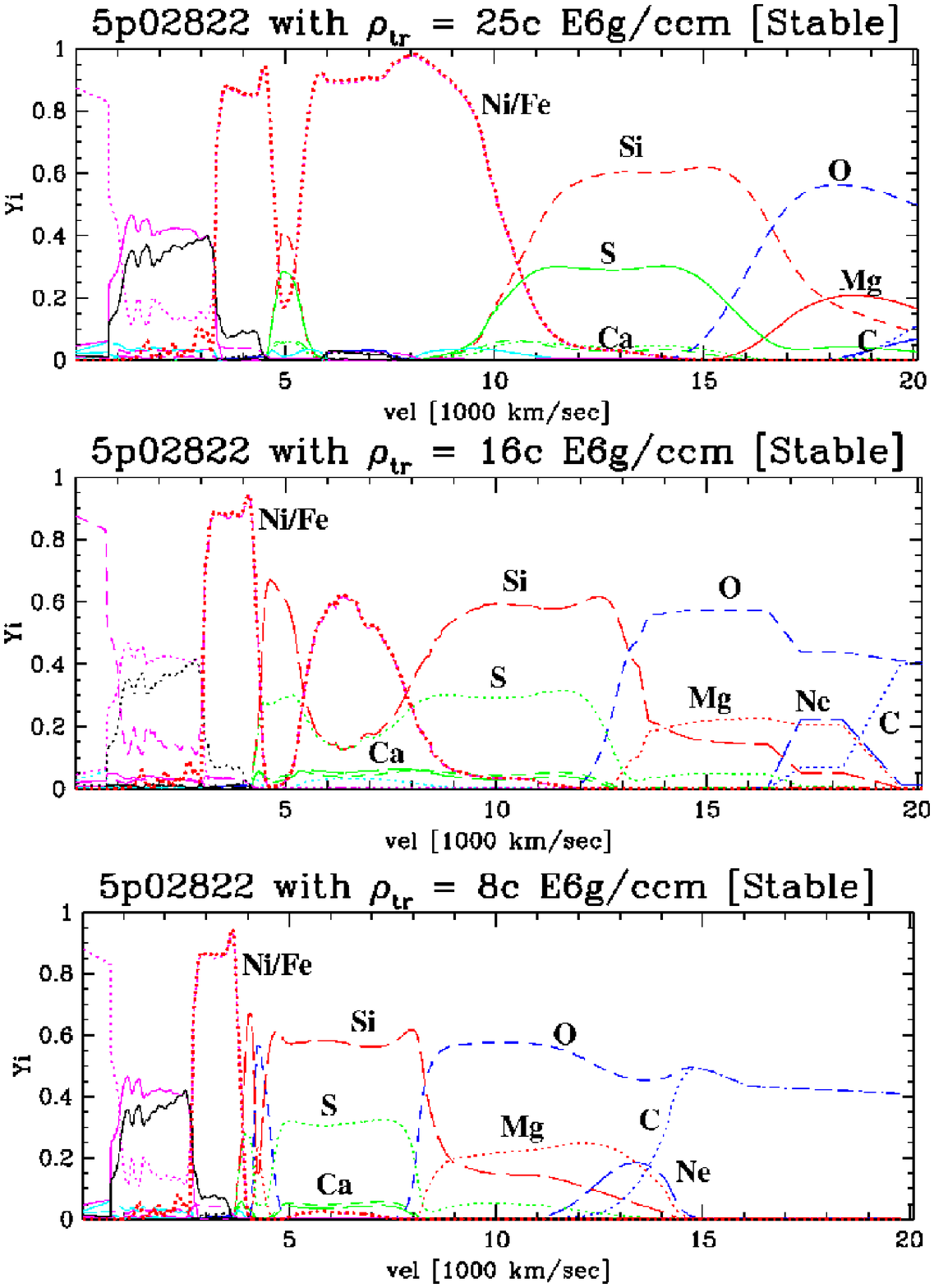}
\vskip -0.0cm
\caption{ Density (blue, dotted) and 
velocity (red, solid) as a function of the mass coordinate [in $M_{Ch}$] (left 
panels), and abundances of
stable isotopes as a function of the expansion velocity (right panels),
 for delayed detonation models with  $\rho_{tr}=$ 8, 16 and 25 $ \times 10^6 
g/cm^3$
(bottom to top).
   All models are based on the same 
$M_{Ch}$ progenitor with a main sequence mass of $3~M_\odot$, solar
metallicity and a central density of $2 \times 10^9 g/cm^3$ at the time of the 
explosion.
These models produce 0.09, 0.26 and 0.6 $M_\odot$ of $^{56}Ni$, respectively.
 In all cases, the entire WD is burned and, thus, all models have 
similar explosion  energies (from \cite{h02}). }
\label{99den} 
\end{figure} 

\begin{figure}[!h] 
\includegraphics[width=6.2cm,angle=270]{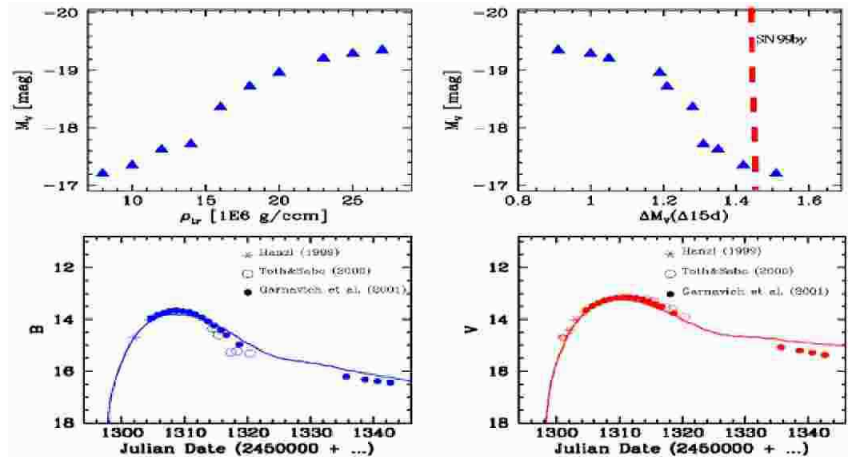}
\vskip -0.0cm
\caption{Maximum brightness $M_V$ as a function of 
$\rho_{tr}$ (upper left) and
$M_V(\Delta M_{\Delta t=15d})$ (upper right) for 
delayed detonation models with $\rho_{tr}$ of 8, 10,
12, 14, 16, 18, 20, 23, 25 and 27~$\times 10^{6}g/cm^3$ from left to right.
 The red, vertical bar (upper right) gives the brightness decline ratio as 
observed for SN1999by.
In the lower panels, the comparison between theoretical and observed B and V 
LCs is  given, implying a distance of $11\pm 2.5$ Mpc, consistent with
independent 
estimates (\cite{bonanos99}). By varying a single parameter, the
transition density at which detonation occurs, a set of models has been 
constructed which spans the observed brightness variation of SNe~Ia.
The absolute maximum brightness depends primarily on the $^{56}Ni$ production,
which for DD-models depends mainly on the transition density $\rho_{tr}$ \cite{hgfs02}.
  The 
brightness-decline relation
 $M_V (\Delta M_{\Delta t =15d})$
  observed in normal bright SNe~Ia
is also reproduced in these models (from \cite{h02}).}
\label{99bylc} 
\end{figure}
\begin{figure}[!h]
\includegraphics[width=6.5cm,angle=270]{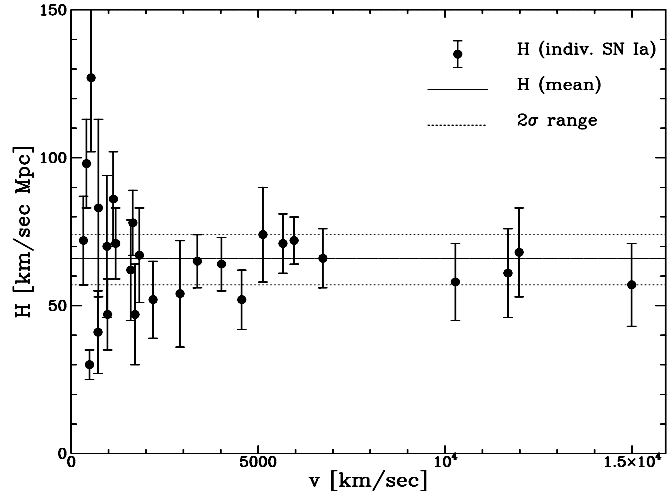}
\caption{
Hubble values H are shown based on model fitting of the
light curves and spectra of 27 individual SNe~Ia including  SN1988U  at a redshift  of 0.38 (not shown)
 \cite{hk96}. We obtain $H_o= 67 \pm 8 km/sec/Mpc$ within a
95 \% level.
 This determination does not depend on $\delta  $-Ceph. calibration or other primary
distance indicators. It is based on basic nuclear physics and spectral constraints,
 and hardly depends  on details of the explosion  models.  One of the main uncertainties is
related to the bolometric correction BC which connects the bolometric luminosity, i.e. the 
of $^{56}Ni $ with the monochromatic brightness. However, the accuracy of the bolometric correction
 can be tested model-independent (see Fig. \ref{bc}).
 The range for $H_o$ owes its stability from spectral constraints. Namely,
 the observed maximum and minimum velocities of the $^{56}Ni $ and Si/S layers which are
$\leq 10,500$ and $\geq 8000 km/sec$ for normal bright SNe~Ia, respectively.
 The former hardens the lower value for $H_o$ because it provides an upper limit for the 
 $^{56}Ni$ mass which can
be speezed within a certain expansion velocity (see \ref{99den}).
 In the same way, the Si/S velocity sets the stage for
the minimum $^{56}Ni$ mass.
}\label{hubble}
\end{figure}
\begin{figure}[!b]
\includegraphics[width=12.0cm,angle=360]{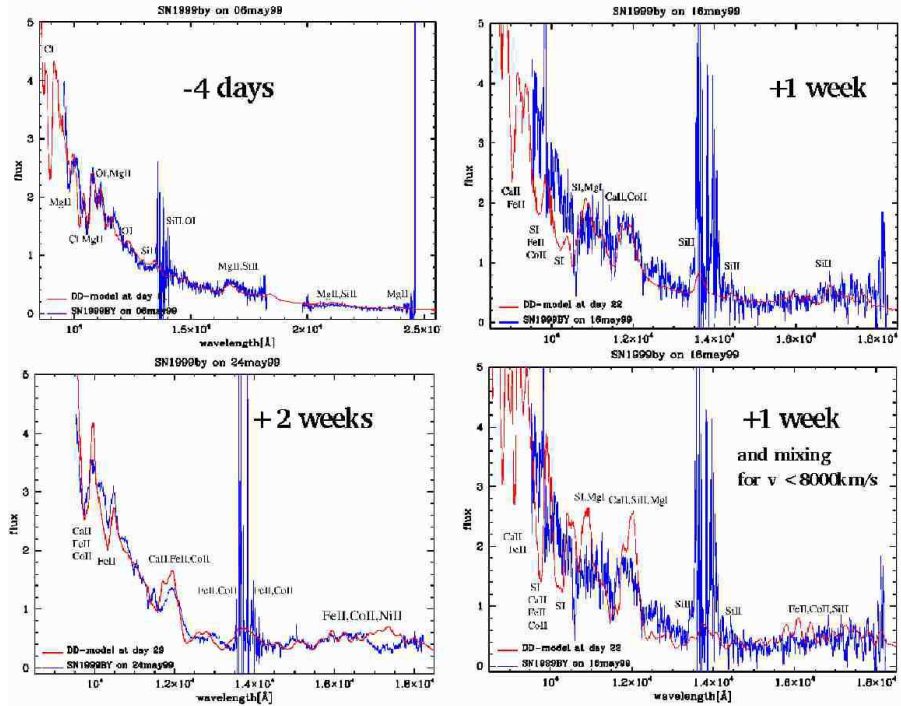}
\vskip -0.0cm
\caption{Comparison of the observed near infrared
spectra of the very subluminous SN99by on May 6 (upper left), May 16 (upper right), and May 24,
1999 (lower left). At those times, the Thomson scattering photosphere is 
located at $v=$ 13,000, 7000  and 4000 $km/s$, respectively. For SN99by, the 
spectra are formed in layers of explosive $C$ and incomplete $Si$ burning
up to about 2 weeks after maximum light. This is in strict contrast to normal 
bright SNe~Ia where the photosphere enters the layers of complete Si burning 
already at about maximum light. In very subluminous SNe~Ia, the transition 
density
is low and the pre-expansion sufficiently large (see Fig. \ref{99den}) that
the layers up to 8000 $km/s$ are not burned to $^{56}Ni$ but to Si only.
Hence, the  $^{56}Ni$ plumes produced during the deflagration phase should 
survive (see Fig. \ref{defl}).
 In the lower right, we show a comparison of the observed and theoretical 
spectrum if  we impose mixing of the inner 0.7 
$M_\odot$.  Obviously, strong mixing of the inner layers can be ruled out (see 
text and \cite{h02}).
 Current 3-D models for the deflagration phase starting from a
static WD are insufficient. Pre-conditioning of the
progenitor is a key element, e.g. turbulent motions in the progenitor or rapid 
rotation.  This is supported by spectropolarimetry of  SN99by which shows an overall
asymmetry  of about 10 \% with a well defined axis \cite{howell01}.}
\label{99byir} 
\end{figure}

\begin{figure}[!h]
\includegraphics[width=3.9cm,angle=270]{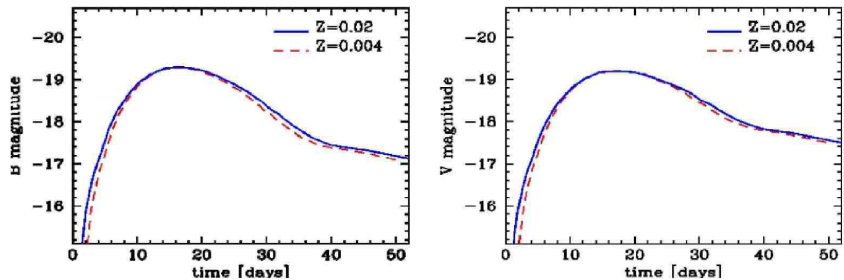}
\caption{Influence of the metallicity Z on the B and V light curves for a 
progenitor star of  7 $M_\odot$ on the main sequence (see \ref{nom}). For the composition 
structure, see Fig.
\ref{nom}.  The explosion model is based on a delayed detonation model typical 
for `normal'  SNe~Ia  (from \cite{hoeflich00}).
 The absolute brightness at 
maximum light is
hardly affected ($\Delta M = 0.02^m$). The rise time changes by about 1d and the 
decline rate over 15 days,  $\Delta M_{\Delta t =15d}$, also changes. When using 
the standard
brightness decline relation, this would produce an offset by $0.1^m$. The dependencies
can be well understood as a consequence of the  mean C/O ratio and the temperature dependence of
the opacities like the brightness decline relation (see text and
\cite{hwt98}). In our  example, the total $^{56}Ni$ production is similar.
As usual, the luminosity at maximum light is provided by both energy due to
instant radioactive decay and  thermal energy stored in the optically thick regions produced 
by  radioactive decays at earlier times.
The total  explosion energies declines with the mean  C/O ratio. The lower expansion rate causes
less energy loss of thermal energy  due to adiabatic expansion
At maximum light,
 the  distance of a given mass element doubles on time scales of $\approx $ 10 to 11  days, respectively.
 At the same time
after the explosion, more energy is
available for low C/O ratios, and the corresponding model shows a slower rise similar to a model with
a larger $^{56}Ni$
production. However, this delay also causes a larger excess of luminosity compared to
the instant energy production by radioactive decays, and the photosphere is slightly cooler.
 The larger excess means a larger total decline past maximum and the cooler photosphere results
a faster receeding of the photosphere. Consequently, the
the decline rate is faster very similar to a slightly less luminous SNe~Ia.
 Models with a lower mean C/O
ratio show a slower increase and and a faster decline \cite{hwt98}.
 Because the limited dependence of the nuclear energy production on the C/O ratio, off-sets in the
brightness decline relation are limited to $\approx 0.3^m$ for the entire range of potential progenitor masses
and metallicities \cite{dominguez01}, and may cause a spread of a similar order around the mean
brightness decline relation.}
\label{nom2}
\end{figure}

\begin{figure}[!h]
\includegraphics[width=4.2cm,angle=270]{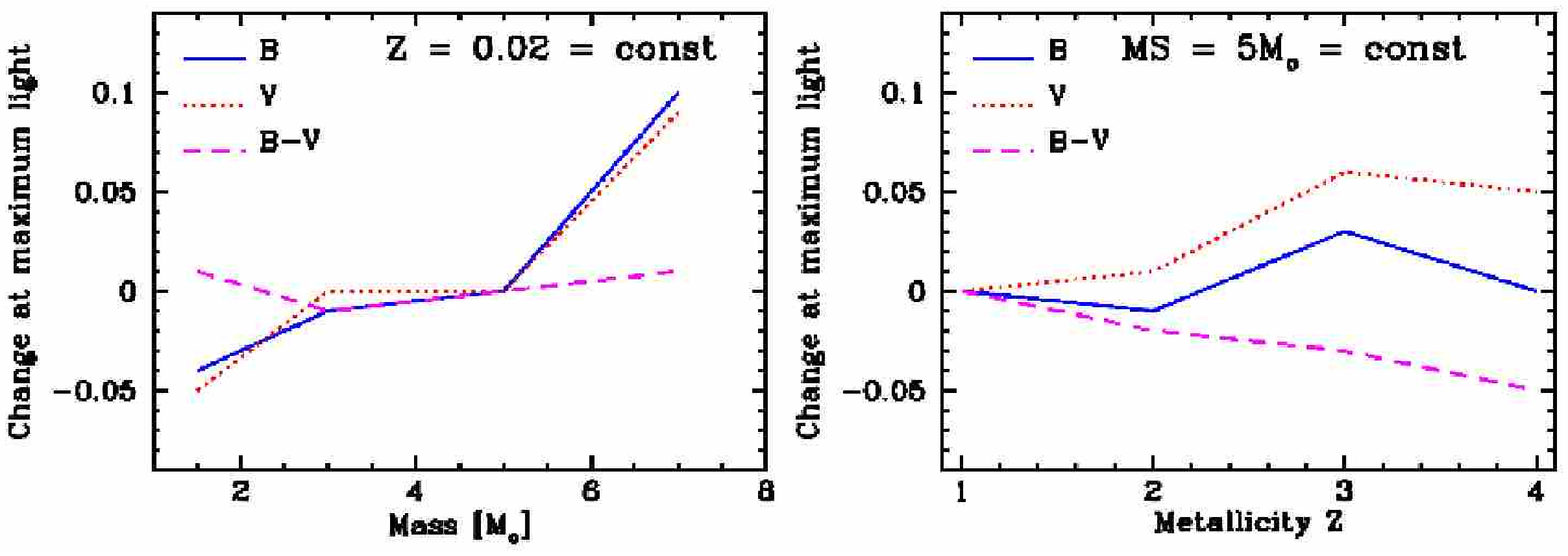}
\vskip -0.0cm
\caption{Influence of  main sequence mass (left) and initial
metallicity (right) on B (---), V ($^{....}$) and B-V (- - -) magnitudes at 
maximum light.
All quantities are given relative to the reference model with a main sequence 
mass of $5 M_\odot$ and solar metallicity. In the right panel, the numbers
1,2,3,4 on the axis refer to Z of 0.02, 0.001, 0.0001 and $10^{-10}$, 
respectively (from
\cite{dominguez01}).
} 
\label{bmv} 
\end{figure}

However, new, high quality data and advances in supernovae simulations 
have opened up new opportunities to constrain the physics of supernovae and to 
improve
the accuracy of their use as standardized candles below  the $0.2^m$ level.
For the first time, a direct relation with the progenitors seems to be 
within reach. In particular, there is mounting evidence that the
properties of the progenitor are directly responsible for the variety 
in
SNe~Ia. These properties include the chemical structure, rotation, and central 
density
(\cite{hwt98}, \cite{wang98}, \cite{hgfs02}).

There is general agreement that SNe~Ia result from some process of combustion 
of a
degenerate WD \cite{hoyle60}.
WDs are the final stages of stellar evolution for all stars with less 
than 7-8 $M_\odot$ (see Fig.~\ref{nom}). During the stellar evolution on the 
main sequence, 
stars gain their energy from central burning of hydrogen to helium 
until H is exhausted in the central region. Subsequently, the star burns 
He to C and O in the center, surrounded by a hydrogen burning shell. 
When He becomes more depleted, the triple-$\alpha$ process becomes less 
efficient and $^{12}C(\alpha,\gamma)^{16}O$ takes over, resulting in an 
inner region of low C abundance (see Fig.~\ref{nom}, right panel). The size of 
the 
He burning core depends on the mass of the star and on 
the metallicity/opacity because it is convective, i.e.~material from 
different radii mixes (e.g.~\cite{dominguez01}).  At these final
stages, the star loses most of its 
mass but with the C/O core remaining: a WD is born. If the star is a 
member of a close binary system it may gain mass at a sufficient rate to
become a SNe~Ia \cite{nomoto82}. Because about 0.2 to 0.7 $M_\odot$
are accreted from an accretion disk, the resulting WD may be 
strongly differential rotating \cite{langer02}.

The exact method of gaining mass sufficient to cause a supernova defines 
three classes of models: (1) An explosion of a CO-WD, with mass close to 
the Chandrasekhar mass $M_{Ch}$, having accreted mass through gravitational 
stripping of the outer layers (called Roche-lobe overflow) 
from an evolved companion star \cite{whelan73}. The explosion is 
mainly triggered by compressional heating near the WD center.  
(2) An explosion 
of a rotating configuration formed from the merging of two low-mass WDs, 
caused by the loss of angular momentum due to gravitational radiation 
from the binary system \cite{webbink84,iben84,paczynski85}.  (3) An 
explosion of a low mass CO-WD triggered by the detonation of a helium 
layer accreted from a close companion \cite{nomoto80,woosley80,woosley86}. 
This third class, the so-called edge-lit sub-Chandrasekhar WD model, has 
been ruled out on the basis of predicted light curves and spectra 
\cite{hoeflich96b,nugent97}. The first model, accretion to $M_{Ch}$, 
is the most successful when compared to observations. 

Within the $M_{Ch}$ scenario (see Fig. \ref{bin}), the free model parameters 
are: 1) The chemical
structure
of the exploding WD -- given by the evolution of the progenitor star 
and the central He-burning; 2) Its central density $\rho_c$ at the 
time of the explosion -- dependent mainly on the accretion rate onto 
the WD; 3) The description of the initial, subsonic burning 
front (deflagration); and 4) The amount of burning prior to the transition from  
deflagration to  detonation (see Fig.  \ref{sce}).
From these, the light curves and evolution of spectra follow directly. Comparison
with observations allow to constrain the parameters for a particular SNe~Ia, its distance and
 the interstellar reddening (see below and Fig. \ref{hubble}).

The first two parameters set the stage.  For the merging scenario, the front 
will start as a detonation making parameters  3 and 4 dependent but adding
the mass of the orbiting envelope as a free parameter.  The $M_{Ch}$ scenario 
requires
parameters 3 and 4 because 
if the WD exploded purely from a thermonuclear runaway reaction then 
almost all of the material would 
burn to $^{56}Ni$, in contradiction to observations that show only about 
0.6 $M_\odot$ is produced.  Instead, a pre-expansion is needed 
to lower the density (see Fig. \ref{sce}).
  This likely occurs during 
an initial phase of a slow deflagration that preserves the 
structure but decreases the binding energy. The lift in potential energy 
depends mainly on the amount of burning, i.e.~total energy produced, 
and almost not at all on the actual rate \cite{dominguez00}. Thus, 
fortunately, details of nuclear burning in the non-linear regime of 
deflagration, about which our understanding is currently limited, will 
hardly affect the final LCs and spectra. 

Successful models need either a rapidly increasing deflagration speed and no 
radial mixing (e.g. W7 \cite{nomoto84} -- see footnote
\footnote{The pure deflagration model W7 is a spherical model and, consequently, 
shows a layered structure which is typical for detonations. Realistic  3-D 
deflagration models do not show a radial layering of the abundances.}),  or a
deflagration-detonation transition (DDT)
(see Fig. \ref{sce}). The detonation or a very rapid deflagration is required to 
match
observations that almost the entire WD is burned (see Fig.
\ref{99den}). Current infrared observations place  tight upper
limits on the amount of unburned material. Depending on the specific SN ~Ia and the quality of
the data, the constraints imposed lie between
 $0.01 $ to  $0.2 M_\odot$ \cite{wheeler98,hgfs02,marion02,rudy02}.
 For general comparison of models, see Fig. \ref{smiley}.
Delayed detonation (DD) models \cite{khokhlov91,woosley94,yamaoka92}, those 
possessing a DDT,  
have been found to reproduce the optical and infrared light curves 
and spectra of ``typical" SNe~Ia reasonably well
\cite{hoeflich95,hkw95,hk96,fisher98,nugent97,wheeler98,lentz01}.
Here the burning starts as a well subsonic deflagration and then turns to a
nearly sonic, detonative mode of burning. Due to the one-dimensional nature of 
the model, the speed of the subsonic deflagration and the moment of the 
transition 
to a detonation are free parameters  hence numbers 3 and 4 mentioned above. 
The moment of deflagration-to-detonation
transition  is conveniently parameterized by introducing the transition
density, $\rho_{\rm tr}$, at which it occurs. 
 The amount of $^{56}$Ni, $M_{56Ni}$, depends primarily on
$\rho_{tr}$ \cite{hoeflich95,hkw95,umeda99},
 and to a much lesser extent on the assumed value  of the deflagration speed,
initial central density of the WD, and initial chemical composition (ratio of
C to O).
In essence this fixes the power source for the supernova light: models 
with a smaller transition density give less nickel and hence both lower 
peak luminosity and lower temperatures \cite{hoeflich95,hkw95,umeda99} 
(Figs.~\ref{99den}, \ref{99bylc}).  This is the first element in
explaining the homogeneity of SNe~Ia. 

The second element is that, in DDs, almost the entire WD is burned, i.e. the 
total production of
nuclear energy is almost constant, and the density and velocity  structures
hardly vary with the $^{56}Ni$ production (Fig. \ref{99den}).
  Together these form the basis of why, to 
first approximation, the SNe~Ia relation between peak magnitude and light 
curve width forms a
one-parameter family. This can be well understood as an opacity 
effect \cite{hoeflich96a}, i.e.~as a consequence of the rapidly dropping
opacity at low temperatures \cite{hoeflich93,khokhlov93}.  Less Ni means
lower temperature so the emitted flux is shifted from the UV towards 
longer wavelengths where there is less line blocking; as a consequence, 
the mean opacities are reduced. Less opacity means the photosphere 
retreats more rapidly to deeper layers, causing a faster release of 
the stored energy and, as a consequence, steeper declining LCs together 
with the decreasing brightness. DD models thus give a natural and 
physically well-motivated origin for the magnitude-light curve width 
relation of SNe~Ia within the paradigm of thermonuclear combustion of 
Chandrasekhar-mass C/O-WDs. These models are able to reproduce light curves and spectra,
and to determine the Hubble constant independently from primary distance indicators (see Fig. \ref{hubble}).
 Furthermore, they can explain
both normal bright and very subluminous SNe~Ia within the same model (Figs. 
\ref{99bylc} and \ref{99byir}).

One of the uncertainties within SN modeling
is the description of the nuclear burning fronts. 
While the propagation of a detonation front is well understood the 
description of the deflagration front and the deflagration to detonation 
transition pose problems. On a microscopic scale, a deflagration 
propagates due to heat conduction by electrons. Though the laminar flame 
speed in SNe~Ia is well known, the front has been found to be 
Rayleigh-Taylor (R-T) unstable (see Fig. \ref{defl})  increasing the effective 
speed of the
burning front \cite{nomoto76}. More recently, significant progress has 
been made toward  a better understanding of the physics of flames.  
Starting from static WDs, hydrodynamic calculations of the deflagration 
fronts have been performed in 2-D \cite{rein99,lisewski00} and 3-D 
\cite{livne93,khokhlov95,khokhlov01}.  It has been demonstrated that 
R-T instabilities govern the morphology of the burning front in the 
regime of linear instabilities, i.e.~as long as perturbations remain 
small.  During the first second after the thermonuclear runaway, the 
increase of the flame surface due to R-T instability remains small and the 
effective burning  speed is close to the laminar speed ($\approx 50\, km/s$) if 
the ignition occurs close to the center. Khokhlov \cite{khokhlov01} also 
shows that the effective burning speed is very sensitive to the energy 
release by the fuel, i.e.~the local C/O ratio.  Therefore, the actual 
flame propagation may  depend on the detailed chemical structure of the
progenitor. Moreover, all current experiments are based on static WDs 
and assumed off-center points of ignition. Recent simulations of the 
final phases before the explosion put the validity of these assumptions 
into question \cite{hs02}. 
Despite advances, the mechanism is not well understood which leads to a 
DDT or, alternatively, to a fast deflagration in the non-linear regime 
of instabilities. Possible candidates for the mechanism are, among others, 
the Zel'dovich mechanism, i.e.~mixing of burned and unburned 
material  \cite{khokhlov97}, crossing shock waves produced in the highly
turbulent medium, or shear flows of rising bubbles at low
densities \cite{livne98,livne98b}. An additional way is  related
shear instabilities present in rapidly, differentially rotating WDs.
 Then, as soon as rising plumes enter this region of
instability, they will be disrupted and strong mixed will occur.  As a consequence,
the burning rate will strongly increase which
may cause a DDT. As discussed above, we must
 expect differential rotation in progenitors because a significant
fraction of the progenitor mass has been accreted from a Kepler-disk.
 Currently, none of the proposed mechanisms have been worked out in detail 
 and  shown to work in the environment of SNe~Ia.
However, as a common factor, all these mechanisms will depend on the 
physical conditions prior to the DDT. In the current state of the art 
we cannot predict a priori the distribution of brightness in a SNe
sample because, within the most favored model, $\rho_{tr}$ determines 
the brightness.  Rather we take this as an input parameter and can 
investigate the small deviations from the brightness decline relation.

 Although pure deflagration models are possible, current 3-D models
show properties inconsistent with the observations (Fig. \ref{smiley}).
 Namely, pure deflagration models predict a significant fraction of the C/O
WD remains unburned and a mixture of burned and unburned material at all 
radii/velocities
(e.g. \cite{barbon90,fisher95} and see  Fig. \ref{99byir}).
Constraints from infrared observations provide good
evidence that  the  WD is  almost fully  incinerated in normal bright SNe~Ia 
\cite{wheeler98,rudy02,marion02}. And
 in contrast to pure deflagration models, in DD-models the detonation front 
 erases the chemical structure
left behind by the deflagration
(Fig. \ref{defl}). Note that the ``classical" deflagration model W7
\cite{nomoto84} shows a layered structure similar to DD-models
 because it has been calculated in spherical geometry rather than the
unlayered structure to be expected from 3-D deflagration models.

 In conclusion, the transition to a detonation or (less likely) to a very fast 
deflagration
determines the $^{56}Ni$ production and causes the one-parameter relation
between peak magnitude and LC width.
To a much lesser extent variations of the other parameters lead to some
deviation from perfect homogeneity on the $0.2^m $ level. For example, an 
increase in the central
density increases the 
electron capture close to the center, shifting the nuclear statistical 
equilibrium away from $^{56}Ni$ \cite{hoeflich96a}. 
Empirically, the magnitude-light curve width relation has been well established 
with a rather 
small statistical error $\sigma$ ($0.18^m$ \cite{hamuy96},$0.12^m~$ 
\cite{riess96},
$0.16^m~$\cite{schmidt98}, $0.14^m~$\cite{phillips99}, 
$0.17^m~$\cite{perl99a}).  These correspond to 5-8\% in distance. Note the 
predicted dispersion for DD models 
is somewhat larger than observed but significantly smaller than generic models 
which show a dispersion of
$0.7^m$ \cite{hoeflich96a}.

\begin{figure}[t] 
\includegraphics[width=7.1cm,angle=270]{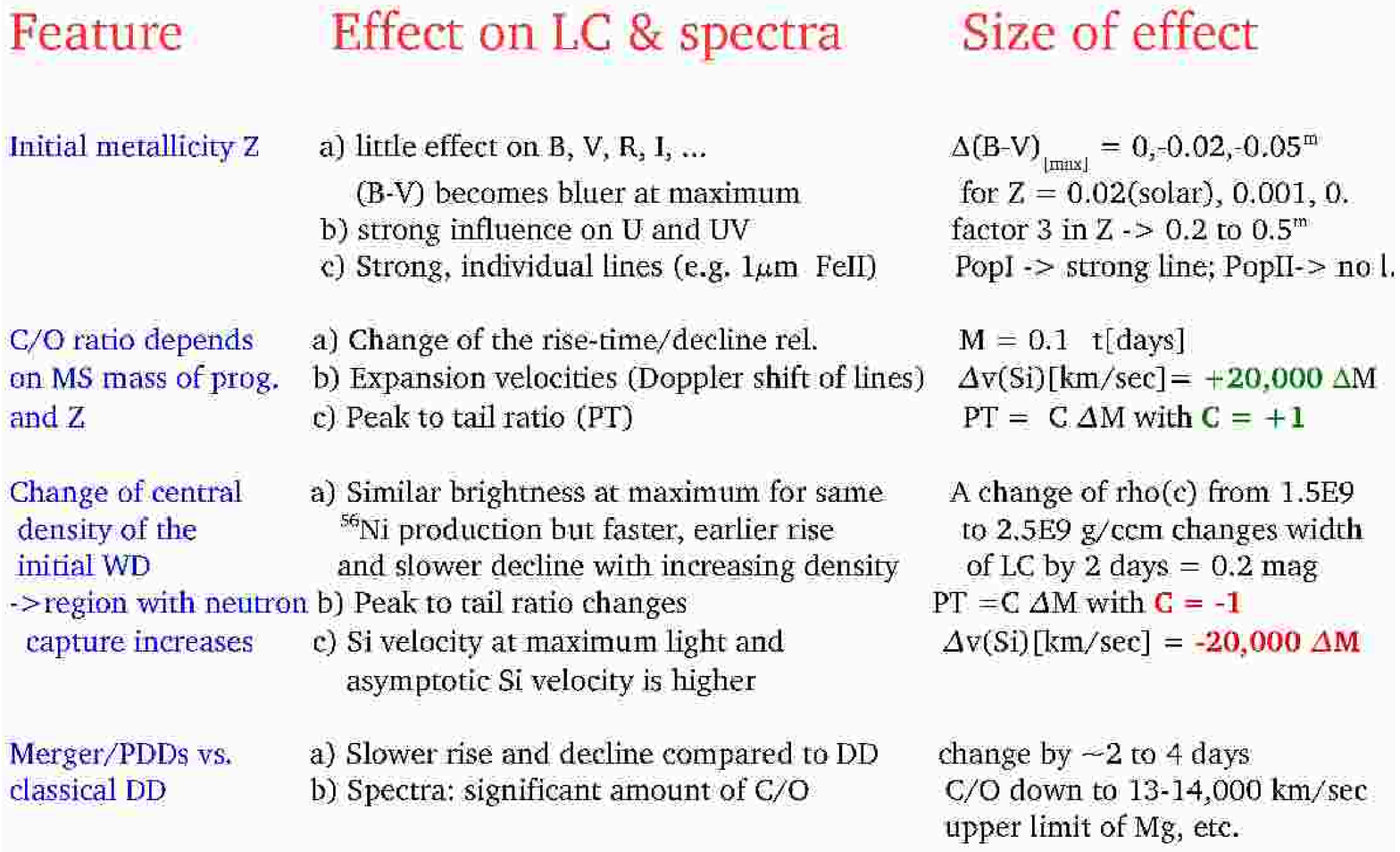}
\caption{Summary of observational effects due to changes in the initial 
metallicity, 
main sequence mass, and central density
for Chandrasekhar mass WD progenitors, and in the
progenitor scenario
\cite{hoeflich95,hk96,hwt98,hoeflich00,dominguez01}.
}
\label{fig2} 
\end{figure}

This may imply a correlation between free model parameters, namely
the properties of the burning front, and the main sequence mass of the 
progenitor $M_{MS}$, metallicity Z, and the central density of the WD at the 
time of the
explosion.  As we have discussed above, there is growing evidence that the final 
outcome
of the explosion is determined by the pre-conditioning  of the WD,
 namely the properties of the WD, its rotation and
the final evolution which leads to the thermonuclear runaway 
\cite{hwt98,hoeflich00,dominguez01,hs02}.
Thus, we must expect that   correlations between observables exist and  can be 
used to further tighten the dispersion
caused by second order parameters (Figs. \ref{nom2} \& \ref{bmv}; see the 
summary table in Fig.
\ref{fig2}).
 From this very argument, we must also expect a shift with redshift in the mean 
properties
of the order of $0.2^m$ due to the population drift in the progenitor 
characteristics and environments.
Nevertheless, the spread around the brightness decline relation may show little
change. As one possible example, the mean metallicity and typical progenitor 
mass at the main
sequence will decrease  with redshift and cause systematic changes in the 
brightness decline relation
(Figs. \ref{nom2} \& \ref{bmv}). These effects can be recognized and compensated 
for if  well observed light curves and spectra are obtained.  Fig. \ref{fig2} 
summarizes many of
the relevant features, their expected size, and their effect on the observables
based on advanced models.
 Note that detailed analyzes of observed spectra and light curves indicate that
mergers and deflagration models such as W7 may contribute to the SN
population \cite{hk96,hatano00}.
 To determine the nature
of the dark energy through the use of SNe~Ia as precision distance indicators,
we need to reduce the residual systematic uncertainties well below the 
statistical dispersions \cite{perl99a,perl99b,albrect00}.
 This is the reason why we need comprehensive observational programs producing
well characterized samples,
from  ground based supernovae surveys such as the Nearby SN Factory \cite{snf}, 
the ESSENCE project 
\cite{essence}, and  the CFH Legacy Survey \cite{cfh} for low redshifts, and the w-project and
space-based missions 
such as the  Supernova/Acceleration Probe \cite{snap} for high redshifts.

\section{Conclusions}
 Supernovae studies have greatly progressed over the last several
years due to advances in both observations and modeling. We are now
able to analyze the explosion and resulting SNe properties in
some detail, and can obtain answers to a number of long standing,
interesting questions.
While we cannot predict a priori the peak magnitude,
 we have seen that we can understand the origin of both the near 
homogeneity and those tight observable relations 
describing the first order deviations. Stability of  the SNe~Ia observables -- stellar amnesia --
 arise
because the nuclear physics determines the structure of the white
dwarfs, and the explosion.
 Although pathways to SNe~Ia span a variety,
the information about the specific history is largely lost along the way to the
progenitor and during the explosion. Thus, convergence due to physics leads 
to a generic accuracy of SNe~Ia as distance
indicators on the $0.2^m$ level. However, these details are needed for
the next level of precision. In particular, pre-conditioning of the
explosion seems to be a key element. These secondary parameters can be revealed 
through detailed models in combination with comprehensive observations which include
both spectra extending to  the near infrared
 and light curves from early times to well after maximum light
  (see table in Fig. \ref{fig2}). This approach is 
 supported by current observations (e.g. \cite{phillips99}).
 Future surveys will provide the rich resource
of data to constrain and refine our understanding of SN progenitors and explosion physics.
 Using all the empirical data the supernovae provide, together 
with the tight relation between the observables and models enables us to 
significantly deduce their  absolute magnitude with confidence.  Thus, SNe~Ia
 are simple, and will be well understood, standardizable
candles for cosmological distance tests. 

\bigskip
\noindent{\bf Acknowledgements:} It is pleasure to thank A. Khokhlov
for the permission to show in Fig. \ref{defl} some of the results of
his 3-D calculations, and for many helpful discussions. This research is supported by
the NASA grant NAG5-7937.

\appendix

\section*{Appendix A: Numerical Radiation Hydrodynamics}

\begin{figure}[ht]
\includegraphics[width=8.7cm,angle=270]{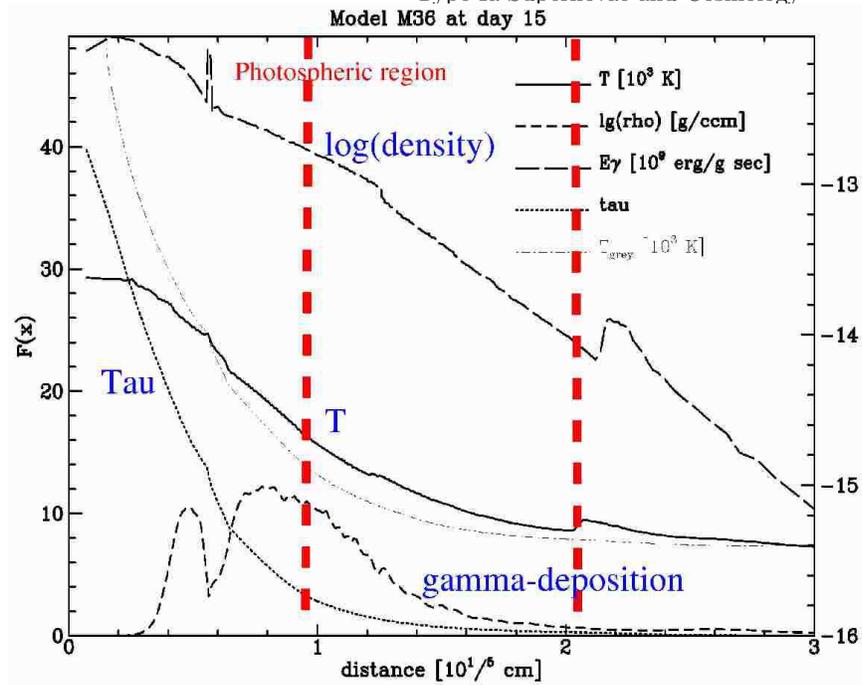}
\vskip -0.04cm
\caption{
Temperature T,  energy deposition due to radioactive
decay $E_{\gamma} $, Rosseland optical depth $Tau$ (left scale) and density log($\rho $) (right scale) are given as a function of distance
 (in $10^{15}cm$) for a typical SNe~Ia at
 15 days after the explosion. For comparison, we give the temperature
$T_{grey}$ for the grey extended atmosphere.
 The light curves and spectra of SNe~Ia are powered by energy release
due to radioactive decay of $^{56}Ni \rightarrow ^{56}Co \rightarrow ^{56}Fe$. The
two dotted, vertical lines indicate the region of spectra formation. 
Most of the energy is deposited within the photosphere and, due to the small optical depth
and densities, strong NLTE effects occur up to the very central region. At maximum light,
the diffusion time scales are comparable to the expansion time scales mandating a consistent
treatment of LCs and spectra
(from H\"oflich 1995).
}
\label{sn94d}
\end{figure}

The computational tools summarized below were used to carry out many of
the analyzes of SNIa and Core Collapse Supernovae ({\cite{H88}, H\"oflich, M\"uller \& Khokhlov 1993,
\cite{hoeflich95}, \cite{howell01}, ...}). A consistent treatment of the explosion, light curves and
spectra are needed (see Fig. \ref{sn94d}).
Details of the numerical methods and codes, namely HYDRA, can be found in 
\cite{h02}  and references therein.  Here we give a brief outline of the basic concepts. 

\begin{figure}[ht]
\hskip 0.0cm \includegraphics[width=9.2cm,angle=270]{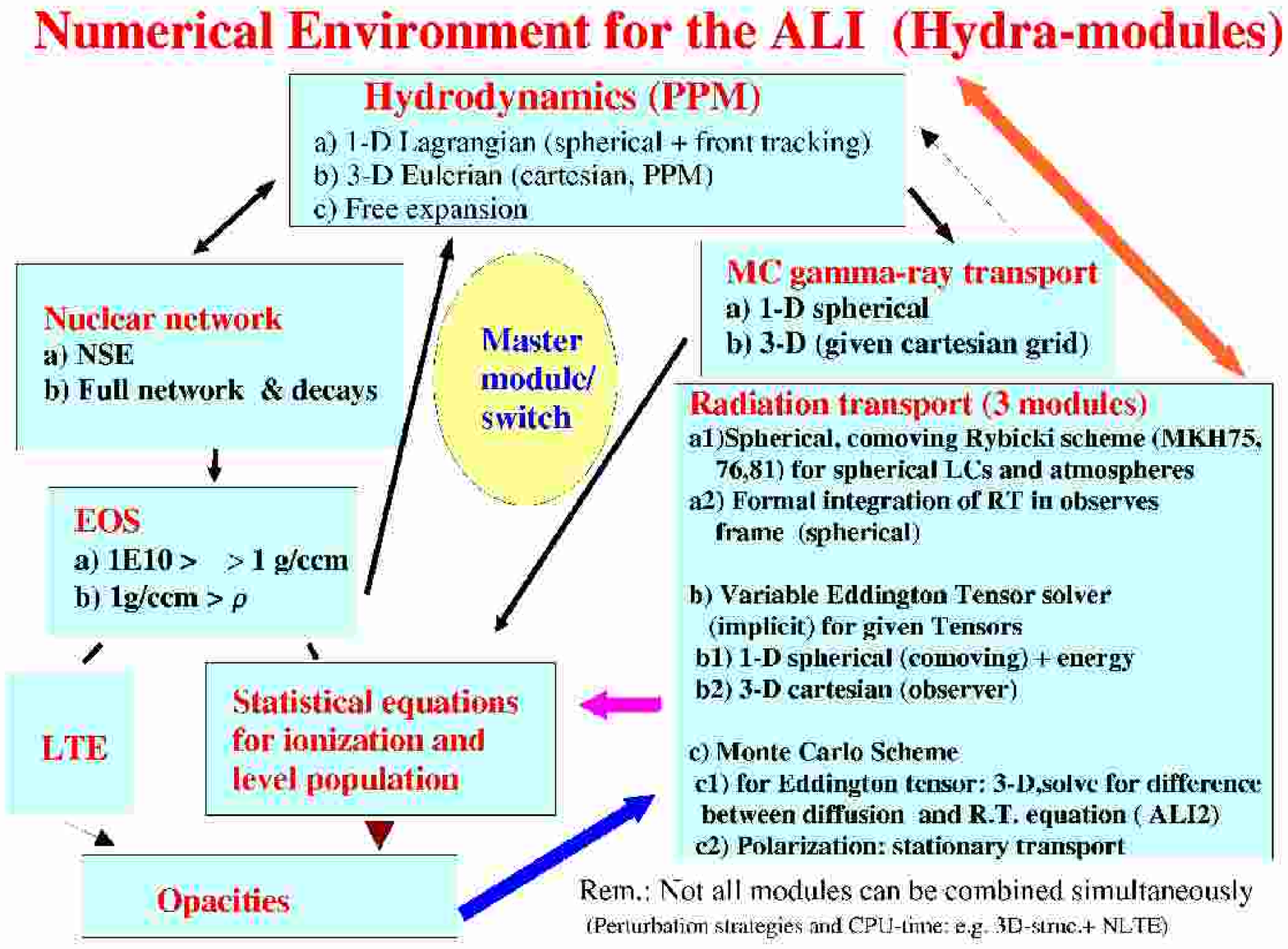}
\caption{
Block diagram of our numerical scheme to solve
radiation hydrodynamical problems including detailed equation of 
state, nuclear and atomic networks. For specific problems, a subset
of the modules is employed  (see text, and e.g. Figs. \ref{99bylc} \& \ref{99byir}).
}
\label{module}
\end{figure}

%

\subsection*{A.1 Hydrodynamics}

The explosions are calculated using a spherical radiation-hydro 
code, including nuclear networks (\cite{hwt98} and
references therein). This code solves the hydrodynamical equations 
explicitly by the piecewise parabolic method \cite{collela84}
and includes the solution of the frequency averaged radiation transport 
implicitly via moment equations, expansion opacities (see below), and a 
detailed equation of state.  Nuclear burning is taken into account using 
a network which has been tested in many explosive environments (see 
\cite{thieleman96}, and references therein). 
The propagation of the nuclear 
burning front is given by the velocity of sound behind the burning front 
in the case of a detonation wave, and in a parameterized form during the 
deflagration phase, calibrated by detailed 3-D calculations 
(e.g.~\cite{khokhlov01}). The density for the transition from deflagration 
to detonation is treated as a free parameter.

\subsection{A.2 Light Curves}

From these explosion models the subsequent expansion and bolometric and 
broad band light curves (LC) are calculated following the method described 
by \cite{hwt98}, and references therein. The LC-code is the same as used 
for the explosion except that $\gamma$-ray transport is included via a 
Monte Carlo scheme and nuclear burning is neglected. In order to allow a 
more consistent treatment of the expansion, we solve the time-dependent, 
frequency-averaged radiation moment equations. The frequency-averaged 
variable Eddington factors and mean opacities are calculated from the 
frequency-dependent transport equations in a co-moving frame at each 
time step.  The averaged opacities have been calculated assuming local 
thermodynamic equilibrium (LTE).  Both the monochromatic and mean 
opacities are calculated in the narrow line limit. Scattering, photon 
redistribution, and thermalization terms, calibrated by the full 
non-LTE-atomic models, have been included. About one thousand frequencies
(in one hundred frequency groups) and about nine hundred depth points 
are used.

\subsection {A.3 Spectral Calculations}

Our non-LTE code (\cite{hoeflich95}, and references therein) solves the 
relativistic radiation transport equations in a co-moving frame.  The 
spectra are computed for various epochs using the chemical, density, 
and luminosity structure and $\gamma$-ray deposition resulting from 
the light curve coder. This provides a tight coupling between the 
explosion model and the radiative transfer.  The effects of 
instantaneous energy deposition by $\gamma$-rays, the stored energy 
(in the thermal bath and in ionization) and the energy loss due to the 
adiabatic expansion are taken into account. Bound-bound, bound-free and 
free-free opacities are included in the radiation transport, which has been 
discretized with about $2 \times 10^4$ frequencies and 97 radial points. 

The radiation transport equations are solved consistently with the 
statistical equations and ionization due to $\gamma$-radiation for the 
most important elements and ions. Typically, between 27 and 137 bound 
levels are used for C, O, Mg, Si, Ca, Ti, Fe, Co, Ni with a total of 
about 40,000 individual NLTE-lines. The neighboring ionization stages 
have been approximated by simplified atomic models restricted to a few 
NLTE levels + LTE levels. The energy levels and cross sections of 
bound-bound transitions are taken from \cite{kurucz93,kurucz94} starting at the
ground state. The bound-free cross sections are taken from TOPBASE 
\cite{mendoza93}. Collisional transitions are 
treated in the `classical' hydrogen-like approximation \cite{mihalas78} 
that relates the radiative to the collisional gf-values. All form factors 
are set to 1. About 10$^{6}$ additional lines are included (out of a line 
list of $4 \times 10^7$) assuming LTE-level populations. The scattering, 
photon redistribution, and thermalization terms are computed with an 
equivalent-two-level formalism.

\section{Appendix B: Uncertainties}

\begin{figure}[!ht]
\includegraphics[width=11.9cm,angle=360]{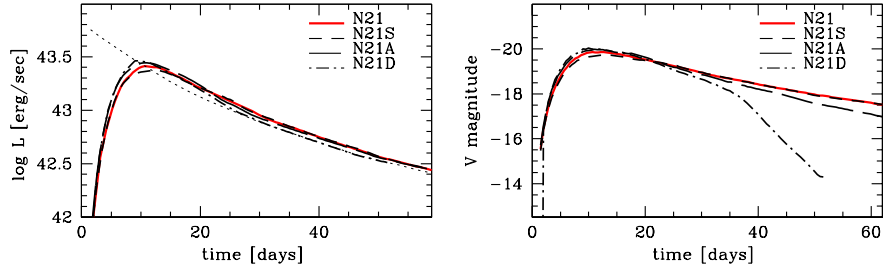}
\vskip -0.0cm
\caption{Systematic study of the influence of physical approximations for an 
early ($\approx 1990$) DD-model based on
frequency averaged LC calculations  \cite{hoeflich93}.
N21: scattering + absorption +
full RT, N21S: N21 but pure scattering lines,  N21A: N21 - but pure absorption 
lines, N21D: diffusion approximation. Nowadays, we use multi-group, NLTE-LCs, 
and more 
realistic WDs are used  (e.g. \cite{hk96,hwt98}).}
\label{freqavg}
\end{figure}

\begin{figure}[!ht]
\includegraphics[width=8.9cm,angle=270]{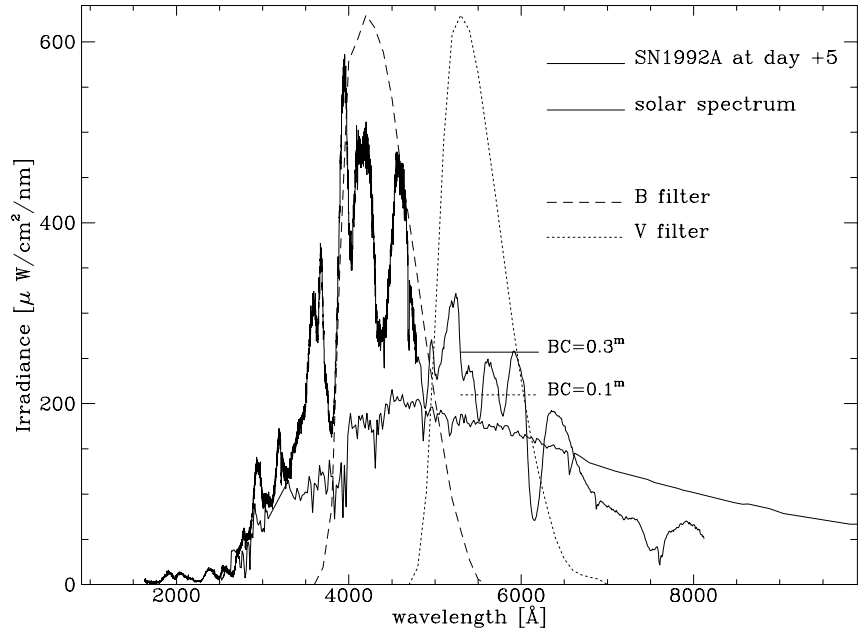}
\vskip -0.0cm
\caption{Observational test for the bolometric correction BC. BC is defined by 
the difference of
a spectral distribution in V compared to the solar irradiance. We give the 
comparison of the solar
flux (thin line) and SN1992A at about 5 days past maximum light 
\cite{kirshner93}.
 In addition, the filter functions for B and V are shown. The horizontal line at about 5500 
\AA ~ (labeled BC=$0.1^m$)
gives the level predicted by the model for SN1999A (from \cite{hk96}). 
 BC=$0.7^m$ and $-0.6^m$  would be required for  values of  $H_o $ being  50 and 80 $ km/sec/Mpc$ 
 and, clearly, can be ruled out. }
\label{bc}
\end{figure}

In the detailed numerical models described in the last section we can 
identify three kinds of uncertainties: 1) uncertainties in the nuclear 
and atomic data such as cross-sections and opacities, 2) errors due to 
inconsistencies, 
discretization, and approximations for the numerical solution, and 3) 
conceptual simplifications in the supernova scenario. 
In Fig. \ref{freqavg} and \cite{hoeflich91,hoeflich93,khokhlov93}, the effects 
of the opacities,
scattering ratio, approximations for (gray) radiation transport and 
different frequency averaging procedures have been tested with respect to 
typical properties of the LCs such as the absolute brightness $M_V$, rise 
time $t_V$ and color index B-V (Figs.~\ref{freqavg}, \ref{bc}).
Before these papers, theoretical models were based on the diffusion 
approximation and opacities that were constant with density, chemistry 
and time.  Clearly, those assumptions were not adequate.  However, within 
reasonable simplifications, the uncertainties in bolometric luminosity 
$L_{bol}$, absolute magnitude in V band $M_V$, time of peak V magnitude 
$t_V$, and color (flux ratio) B-V have been found to be less than 10\%,
even if the opacities have been scaled by a factor of 3 either way. 

Within this narrow range the numerical solution of the radiation transport 
problem helps to improve the differences between $L_{bol}$ and the 
energy production due to $^{56}Ni$.  As discussed in sections III and IV there exists a
strong physical basis for this as well as tight model restrictions on the 
variation. Extensive tests showed
variations in $L_{bol}/E_{\gamma }$ are less than $\pm 20\%$ even if we allow 
for
model assumptions which have  since been ruled out
or for use of clearly simplified approximations (e.g. one-zone model).
The exact time of maximum light is 
governed by when the temperature drops below $\approx 10,000 K$, 
i.e.~when the mean opacity drops by several orders of magnitude 
\cite{hoeflich91,hoeflich93,khokhlov93}. Prior to maximum light, 
the drop in the 
temperature is governed by the expansion work and only to a small degree 
by radiation transport effects \cite{hoeflich93}. Note that V is 
well determined because it is in the linear tail of the emissivity. 
Uncertainties in B, in particular past maximum light, have been found to 
be up to  $0.2^m$ because the size of line blocking and photon 
redistribution effects change drastically over the period considered. 

We can also test the global energy conservation 
based purely on observations and predictions.  Fig.~\ref{bc} shows the 
relation between luminosity and the monochromatic colors, known as 
the bolometric correction BC.  The empirical and model based factors for 
BC agree to better than $0.1^m$ \cite{hk96}. Another class of test
is based on predictions of distances for individual SNe~Ia 
\cite{hoeflich92,mueller94} made prior to their determination based on 
$\delta Ceph$ stars by HST: all but one agreed well within the $1 
\sigma$-error bars (see table 1 from \cite{hk96}). A notable exception 
was the peculiar SN1991T for which \cite{hoeflich91} predicted a distance 
of $14.5\pm2$ Mpc while  distance measurements of a neighboring galaxy
(host of 60f) suggested a distance of 19Mpc \cite{saha97}. However,
recently a direct measurement of the host galaxy by $\delta Ceph$ 
reduced the distance to 13.5 Mpc \cite{saha01}.

Discretization errors have been tested by doubling the number of depth 
points (e.g.~\cite{khokhlov93}, \cite{hoeflich95}).  Typically, we use 456 
to 912 depth points. The errors are found to be less than a few percent 
in the total energy and the production of elements during the explosion. 
For the LCs and spectra, the resulting fluxes change by less than 1\% 
\cite{hoeflich95}. In the LC flux calculations, the main sources of 
errors are due to the limitations of the frequency grid, the neglect of 
aberration terms in the radiation transport equation, and the use of 
simplified atomic models for the frequency redistribution of photons. 
If we compare the LCs with the spectral calculations, the resulting 
error is $<  10\%$ in the flux and about $0.05^m$ and $0.2^m$ in B-V around
maximum light and about 2 weeks after maximum, respectively 
\cite{hoeflich95,bowers97,hwt98,hgfs02}.

Errors can arise because the models are simplifications of reality, 
e.g.~adopting spherical symmetry.  Deviations from this could be due to 
either global asymmetries in the density or the distribution of elements 
\cite{wang97,wang01,howell01}. In general, only upper limits for normal 
bright SNe~Ia are given; recent observations with VLT indicate a level 
of about 0.1\%, which translates into a direction dependent luminosity 
of $\approx 0.1^m$ \cite{hoeflich91}. However, the subluminous SN1999by
shows polarization as high as 0.7\% \cite{howell01}, and rotational 
symmetry. This implies that the luminosity will vary by about $0.3^m$,
depending on the position of the observer. 
 
Another possible breakdown in geometry is the description of the burning 
front but, currently, the size of this  effect is hard to estimate.
As noted above, 3-D  models are currently limited to the
regime of linear instabilities and a significant amount of C/O remains 
unburned ($\approx 0.5-0.8 M_\odot$). Clearly, these early 3-D attempts 
are in contradiction with observations.

\end{document}